\documentclass[aip,reprint]{revtex4-1}

\usepackage{amsmath}
\usepackage{graphicx}
\usepackage{color}
\begin{document}


\title{AC Elastocaloric effect as a probe for thermodynamic signatures of continuous phase transitions} 



\author{M.\,S.\,Ikeda}
\email[email: ]{matikeda@stanford.edu}
\affiliation{\mbox{Geballe Laboratory for Advanced Materials, Stanford University, 476 Lomita Mall, Stanford, CA 94305, USA}}
\affiliation{\mbox{Department of Applied Physics, Stanford University, 348 Via Pueblo Mall, Stanford, CA94305, USA}}
\affiliation{Stanford Institute for Materials and Energy Science, SLAC National Accelerator Laboratory, 2575  Sand  Hill  Road,  Menlo  Park,  California  94025,  USA}

\author{J.\,A.\,W.\,Straquadine}
\affiliation{\mbox{Geballe Laboratory for Advanced Materials, Stanford University, 476 Lomita Mall, Stanford, CA 94305, USA}}
\affiliation{\mbox{Department of Applied Physics, Stanford University, 348 Via Pueblo Mall, Stanford, CA94305, USA}}
\affiliation{Stanford Institute for Materials and Energy Science, SLAC National Accelerator Laboratory, 2575  Sand  Hill  Road,  Menlo  Park,  California  94025,  USA}

\author{A.\,T.\,Hristov}
\affiliation{\mbox{Geballe Laboratory for Advanced Materials, Stanford University, 476 Lomita Mall, Stanford, CA 94305, USA}}
\affiliation{Stanford Institute for Materials and Energy Science, SLAC National Accelerator Laboratory, 2575  Sand  Hill  Road,  Menlo  Park,  California  94025,  USA}
\affiliation{\mbox{Department of Physics, Stanford University, 382 Via Pueblo Mall, Stanford, CA 94305,USA}}

\author{T.\,Worasaran}
\affiliation{\mbox{Geballe Laboratory for Advanced Materials, Stanford University, 476 Lomita Mall, Stanford, CA 94305, USA}}
\affiliation{\mbox{Department of Applied Physics, Stanford University, 348 Via Pueblo Mall, Stanford, CA94305, USA}}
\affiliation{Stanford Institute for Materials and Energy Science, SLAC National Accelerator Laboratory, 2575  Sand  Hill  Road,  Menlo  Park,  California  94025,  USA}

\author{J.\,C.\,Palmstrom}
\affiliation{\mbox{Geballe Laboratory for Advanced Materials, Stanford University, 476 Lomita Mall, Stanford, CA 94305, USA}}
\affiliation{\mbox{Department of Applied Physics, Stanford University, 348 Via Pueblo Mall, Stanford, CA94305, USA}}
\affiliation{Stanford Institute for Materials and Energy Science, SLAC National Accelerator Laboratory, 2575  Sand  Hill  Road,  Menlo  Park,  California  94025,  USA}

\author{M.\,Sorensen}
\affiliation{\mbox{Geballe Laboratory for Advanced Materials, Stanford University, 476 Lomita Mall, Stanford, CA 94305, USA}}
\affiliation{\mbox{Department of Applied Physics, Stanford University, 348 Via Pueblo Mall, Stanford, CA94305, USA}}
\affiliation{Stanford Institute for Materials and Energy Science, SLAC National Accelerator Laboratory, 2575  Sand  Hill  Road,  Menlo  Park,  California  94025,  USA}

\author{P.\,Walmsley}
\affiliation{\mbox{Geballe Laboratory for Advanced Materials, Stanford University, 476 Lomita Mall, Stanford, CA 94305, USA}}
\affiliation{\mbox{Department of Applied Physics, Stanford University, 348 Via Pueblo Mall, Stanford, CA94305, USA}}
\affiliation{Stanford Institute for Materials and Energy Science, SLAC National Accelerator Laboratory, 2575  Sand  Hill  Road,  Menlo  Park,  California  94025,  USA}

\author{I.\,R.\,Fisher}
\affiliation{\mbox{Geballe Laboratory for Advanced Materials, Stanford University, 476 Lomita Mall, Stanford, CA 94305, USA}}
\affiliation{\mbox{Department of Applied Physics, Stanford University, 348 Via Pueblo Mall, Stanford, CA94305, USA}}
\affiliation{Stanford Institute for Materials and Energy Science, SLAC National Accelerator Laboratory, 2575  Sand  Hill  Road,  Menlo  Park,  California  94025,  USA}



\date{\today}

\begin{abstract}
Studying the response of materials to strain can elucidate subtle properties of electronic structure in strongly correlated materials. So far, mostly the relation between strain and resistivity, the so called elastoresistivity, has been investigated. The elastocaloric effect is a second rank tensor quantity describing the relation between entropy and strain. In contrast to the elastoresistivity, the elastocaloric effect is a thermodynamic quantity. Experimentally, elastocaloric effect measurements are demanding since the thermodynamic conditions during the measurement have to be well controlled. Here we present a technique to measure the elastocaloric effect under quasi adiabatic conditions. The technique is based on oscillating strain, which allows for increasing the frequency of the elastocaloric effect  above the thermal relaxation rate of the sample. We apply the technique to Co-doped iron pnictide superconductors and show that the thermodynamic signatures of second order phase transitions in the elastocaloric effect closely follow those observed in calorimetry experiments. In contrast to the heat capacity, the electronic signatures in the elastocaloric effect are measured against a small phononic background even at high temperatures, establishing this technique as a powerful complimentary tool for extracting the entropy landscape proximate to a continuous phase transition.
\end{abstract}


\maketitle 

\section{Introduction}

The elastocaloric effect (ECE) is described by a second-rank tensor that relates the deformation of a material to the resulting change in its temperature under adiabatic conditions. Whilst the ECE has been studied extensively in the context of refrigeration (typically utilizing first-order martensitic phase transformations\cite{Rod80.1,Bon08.1,Qia16.1} to induce cooling), it has been overlooked in the study of quantum materials and their associated phase transitions. It has recently been demonstrated in a range of topical materials that strains are a powerful tool through which to probe and tune novel quantum phases\cite{Hic14.2,Rig15.1,Ike18.1,Kim18.1}. Here we propose the ECE as a sensitive probe of the strain-derivatives of the entropy landscape, focusing in particular on second-order phase transitions, and demonstrate a quasi-adiabatic technique through which to measure the elastocaloric response to an oscillating uniaxial stress. Using the electronic-nematic and magnetic phase transitions in Ba(Fe$_{1-x}$Co$_x$)$_2$As$_2$ as a test case, we demonstrate that the ECE is uniquely sensitive to second order phase transitions among thermal probes.


To date, ECE measurements\cite{Osm14.1} have been performed by quickly ramping the applied force to a material and measuring the temperature response within timescales corresponding to a quasi adiabatic condition. These experiments require high accuracy measurements of rapidly changing signals. The second, indirect approach applied in literature is to iso-thermally measure the elastocaloric entropy change by carefully determining the stress-strain relation of a given material\cite{Bon08.1}. Here we discuss a new, alternative approach in which direct ECE measurements are performed under oscillating uniaxial stress. We show that this technique allows for (i) controlling the adiabatic condition by choosing the appropriate stress frequency and (ii) achieving a large signal-to-noise ratio by capitalizing on standard phase sensitive lock-in techniques. When comparing to AC calorimetry\cite{Sul68.1,Vel92.1,Rio04.1}, the most striking advantage of the technique discussed here is the fact that for the ECE, electronic signatures are measured against a small contribution from lattice degrees of freedom\footnote{A phonon contribution to the elastocaloric effect is expected due to the strain dependence of the sound velocity, which changes the entropy from lattice degrees of freedom at a given temperature. This effect is typically small compared to strain driven entropy changes in the vicinity of continuous phase transitions.} even at high temperatures. In addition, we note that the ECE is an intensive quantity, whereas the heat capacity is extensive. Elastocaloric effect measurements can thus operate in a wide frequency regime, only limited by the time constant of thermalization of the thermometer or the experimentally achievable stress frequencies. 
 
Applying (uniaxial) stress to a material under adiabatic conditions 
\begin{equation}
dS=\left(\frac{\partial S}{\partial T}\right)_{\sigma} dT + \left(\frac{\partial S}{\partial \sigma_{\rm ij}}\right)_{T,\sigma^{\rm 1}}d\sigma_{\rm ij}=0,
\end{equation}
and noting that $(\partial{S}/\partial{T})_\sigma = C_{\rm \sigma}/T$ and $\left(\partial{S}/\partial{\sigma_{\rm ij}}\right)_{T,\sigma^{\rm 1}}$ equals the thermal expansion coefficient $\alpha_{\rm ij}$ ($\sigma^{\rm 1}$ indicates that all stress components are kept constant except for one), an elastocaloric temperature change
\begin{equation}
dT=-\frac{T}{C_{\rm \sigma}}\alpha_{\rm ij}d\sigma_{\rm ij}=-\frac{T}{C_{\rm \sigma}} \alpha_{\rm ij}c_{\rm ijkl}d\epsilon_{\rm kl}
\label{Eq1}
\end{equation}
is expected in response to the applied stress. Here, $T$ is the sample temperature, $c_{\rm ijkl}$ are the components of the elastic stiffness, and $\epsilon_{\rm kl}$ the components of the strain tensor. $C_{\rm \sigma}$ and $\alpha_{\rm ij}$ are the heat capacity and the thermal expansion components, respectively. Both properties assume all components of the stress tensor $\boldsymbol{\sigma}$ to be constant. Summation over equal indices is assumed.

In the vicinity of second order phase transitions the elastocaloric coefficients $\left(dT/d\varepsilon_{\rm xx}\right)_S$ \footnote{Adiabatic conditions are assumed throughout this manuscript. The constant entropy label is thus dropped in all the equations below.} are related to the critical contribution of the heat capacity $C_{\rm \sigma}^{\rm (c)}$ via
\begin{equation}
\frac{dT}{d\varepsilon_{\rm xx}}=\frac{C_{\rm \sigma}^{\rm (c)}}{C_{\rm \sigma}}\frac{dT_{\rm c}}{d\varepsilon_{\rm xx}}+\frac{dT^{\rm (nc)}}{d\varepsilon_{\rm xx}},
\label{ECCP}
\end{equation}
where $dT_{\rm c}/d\varepsilon_{\rm xx}$ describes the strain dependence of the transition temperature $T_{\rm c}$, and $dT^{\rm (nc)}/d\varepsilon_{\rm xx}$ a (typically small and weakly temperature dependent) contribution from non critical degrees of freedom. Note that the total derivative with respect to $\varepsilon_{\rm xx}$ captures that uniaxial stress $\sigma_{\rm xx}$ typically (for Poisson ratios $\nu=-\varepsilon_{rm yy}/\varepsilon_{rm xx}$ and $\nu^\prime-\varepsilon_{rm zz}/\varepsilon_{rm xx}$ not equal to zero) causes strains $\varepsilon_{\rm xx}$, $\varepsilon_{\rm yy}$, and $\varepsilon_{\rm zz}$ (see Appendix\,\ref{sec:thermodyn}).

\begin{figure}
 \includegraphics[width=0.5\textwidth]{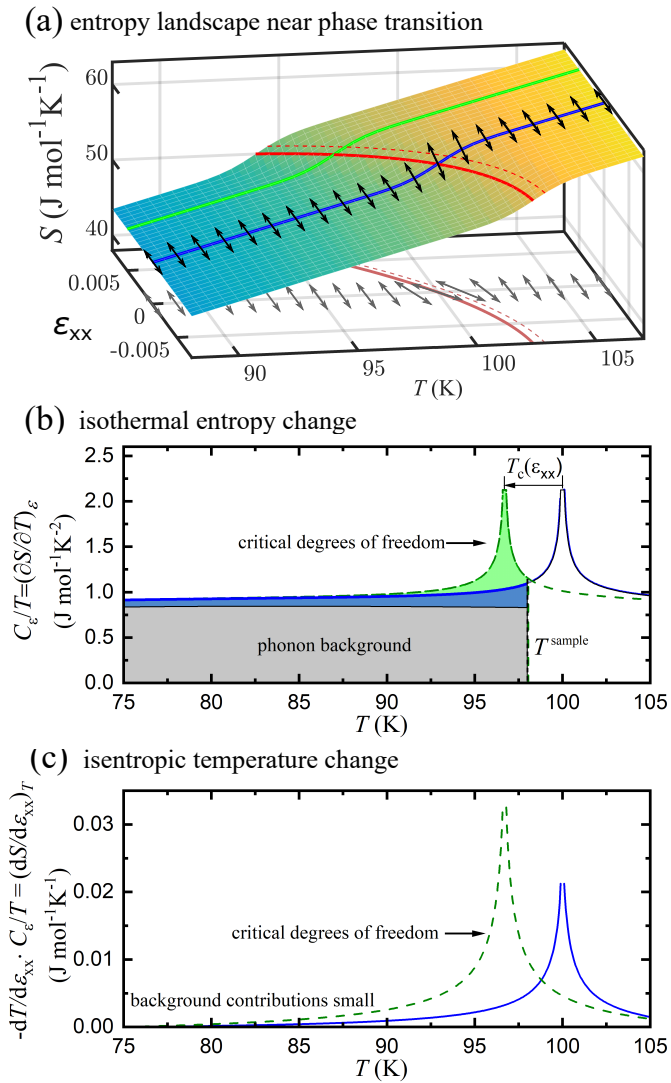}
 \caption{(a) Sketch showing the entropy landscape for a material undergoing a continuous phase transition (with an upper critical dimension larger than the dimension of the appropriate universality class) as a function of temperature and strain. The solid red lines on the entropy surface and the $T$-$\varepsilon_{\rm xx}$-plane mark the phase transition ($T_{\rm c}=100-520\varepsilon_{\rm xx}-28300\varepsilon_{\rm xx}^2$, see Ref.\,\onlinecite{Ike18.1}). The dashed red lines show paths parallel to the transition lines (see discussion in Appendix\,\ref{sec:thermodyn}). The green and blue line on the entropy surface highlight the entropy at constant strains $\varepsilon_{\rm xx}=0$, and $\varepsilon_{\rm xx}=0.005$. The black arrows on the entropy surface sketch AC strain (the strain amplitude within the actual experiment is on the order of 1E-4) probing $d{S}/d{\varepsilon_{\rm xx}}$; the grey arrows in the $T$-$\varepsilon_{\rm xx}$-plane highlight elastocaloric temperature oscillations in the vicinity of the phase transition. Panel (b) shows the partial derivative of the green and blue lines of panel (a) with respect to temperature. This quantity corresponds to the heat capacity at constant volume divided by temperature. In case the sample temperature is kept constant in the vicinity of the critical temperature, an isothermal strain change causes a large entropy change (are colored in green) due to the critical contribution shifting in temperature. Panel (c) shows the derivative of S with respect to strain $\varepsilon_{\rm xx}$ for the DC two offset strains $\varepsilon_{\rm xx}=0$ (solid blue line), and $\varepsilon_{\rm xx}=0.005$ (dashed green line). For isentropic strain changes, this quantity translates into elastocaloric temperature changes $dT/d\varepsilon_{\rm xx}$.}
 \label{Fig0}
\end{figure}

 A detailed derivation of equation \ref{ECCP} is presented in Appendix \ref{sec:thermodyn}. Physical intuition can be gained by considering the entropy of the sample as a function of temperature and strain. Figure\,\ref{Fig0}(a) sketches the entropy landscape of a material undergoing a continuous phase transition at a temperature $T_{\rm c}(\varepsilon_{\rm xx})$ (where for purposes of illustration we assume that $T_{\rm c}=100-520\varepsilon_{\rm xx}-28300\varepsilon_{\rm xx}^2$, which is the strain dependence of the nematic transition observed\cite{Ike18.1} for a material very similar to the sample studied below). The phase transition line is shown on the entropy surface as well as in the temperature-strain plane as a solid red line. The strain dependence of the non-critical degrees of freedom (e.g. the dependence of the sound velocity on strain etc.) are typically very small and are neglected in Fig.\,\ref{Fig0}(a). We now consider two distinct scenarios; (i) isothermal (slow strain changes) and (ii) isentropic (rapid) changes with strain. The blue and the green lines on the entropy surface mark the entropy at constant strains as a function of temperature for different offset strains $\varepsilon_{\rm xx}=0$, and  $\varepsilon_{\rm xx}=0.005$, respectively. Figure\,\ref{Fig0}(b) shows the derivative of these two lines with respect to temperature which corresponds to the heat capacity at constant strain divided by temperature. Note that the difference of the heat capacity at constant strain and stress (see Appendix \ref{sec:thermodyn}, Eqn.\,\ref{dCp}) is small for solids at moderately low temperatures and is neglected in the discussion presented here. As mentioned above, the background contribution is unchanged for both lines. The critical contribution on the other hand moves to lower temperatures under tensile strain which, if the sample is held at a constant temperature $T^{\rm sample}$, increases the entropy [area colored in green in Fig.\,\ref{Fig0}(b)]. If the strain changes are done fast compared to the time scale of thermalization of the sample and its surrounding $\tau_{\rm i}$, the isentropic scenario (ii) has to be considered. In this case, the total entropy change $dS$ is equal to zero and thus the sample temperature will follow the critical temperature to compensate the strain related entropy change. Considering only small changes of strain $d\varepsilon$, the entropy balance can be written as:
 \begin{equation}
     dS = \frac{C_{\rm \sigma}^{\rm (c)}}{T} \frac{dT_{\rm c}}{d\varepsilon}d\varepsilon-\frac{C_{\rm \sigma}}{T} dT=0.
 \end{equation}
 Here, the first term describes the entropy gained from shifting the critical contribution $C_{\rm \sigma}^{\rm (c)}/T$ to lower temperatures and the second term accounts for the balancing contribution due to lowering of the sample temperature. Rearranging this expression, and having negelcted the strain dependence of the non critical degrees of freedom, we arrive at Eq.\,\ref{ECCP}. A more rigorous thermodynamic derivation is discussed in Appendix\,\ref{sec:thermodyn}.
 
 We measure this effect by slowly sweeping temperature while applying fast AC strain  (optionally on top of a DC strain offset). The black arrows drawn onto the entropy landscape and the grey arrows in the $T$-$\varepsilon_{\rm xx}$-plane (Fig.\,\ref{Fig0}(a)) sketch AC strain, probe the derivative $d{S}/d{\varepsilon_{\rm xx}}$. In the vicinity of the critical temperature this quantity is strongly enhanced due to the critical contribution to the entropy. This causes elastocaloric temperature oscillations which cause the grey arrows in the $T$-$\varepsilon_{\rm xx}$-plane to tilt with respect to the $\varepsilon_{\rm xx}$-axis. The magnitude of this effect as a function of temperature is shown in Figure\,\ref{Fig0}(c) for two different DC offset strains $\varepsilon_{\rm xx}=0$, and $\varepsilon_{\rm xx}=0.005$. As stated in Eq.\,\ref{ECCP} the elastocaloric temperature oscillation is proportional to $C_{\rm \sigma}^{\rm (c)}/C_{\rm \sigma}$ multiplied by $dT_{\rm c}/d\varepsilon_{\rm xx}$ (which is larger for $\varepsilon_{\rm xx}=0.005$ in our specific illustration), and is typically measured against a small background [$dT^{\rm (nc)}/d\varepsilon_{\rm xx}$ is small in comparison to $(C_{\rm \sigma}^{\rm (c)}/C_{\rm \sigma})\cdot (dT_{\rm c}/d\varepsilon_{\rm xx})$]. This is the specific advantage that ECE measurements provide over standard calorimetry in terms of determining the critical contribution to the entropy of the system. Generally, the mechanism discussed here should be expected for any second order phase transition tunable by strain.

The key experimental challenge in performing thermal measurements under strain is that in order to apply the strain the sample must be in good mechanical contact with the strain apparatus, and hence also good thermal contact. The adiabatic condition cannot therefore be fulfilled to any reasonable approximation, and so a quasi-adiabatic approach is more appropriate. By applying an oscillating strain with a frequency greater than the inverse relaxation time $\tau_{\rm i}^{-1}$ (that defines the time scale of the heat exchange of the sample and its surrounding) the temperature oscillation in the sample approaches its absolute value in the region of the sample that is well spaced (as characterized by the thermal length) from the mechanical connections. As such, following an introduction to the experimental setup, we discuss in detail the appropriate models through which to understand the frequency response of the sample and sensor which is essential in producing good measurements. It is demonstrated that the quasi-adiabatic condition can be fulfilled over a broad frequency range, but also that the absolute value of the ECE is reduced by the heat capacity of the temperature sensor as frequency increases. Finally, we discuss our results applying this method to the iron-based superconductor Ba(Fe$_{0.979}$Co$_{0.021}$)$_2$As$_2$, and show that AC-ECE is a powerful tool to extract thermodynamic signatures of second order phase transition with a very high accuracy, even at high temperatures.

\section{Experimental Technique}
\subsection{Experimental setup}
\begin{figure}
 \includegraphics[width=0.5\textwidth]{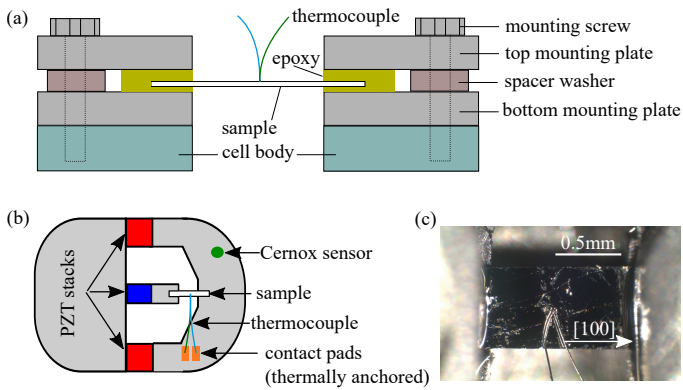}
 \caption{(a) Sketch showing the cross sectional side view of a sample mounted on a CS100 Razorbill cell. The sample is attached to the top and bottom mounting plates using Devcon 2-ton epoxy. Details on the mounting technique are reported elsewhere \cite{Ike18.1}. For the elastocaloric effect measurements discussed here, the sample is contacted using a type-E thermocouple. The thermocouple is thermally referenced to the titanium cell by using well heat sunk contact pads attached to the cell body. (b) Schematic diagram showing the top view of our experiment before mounting the top mounting plates. The outer red PZT stacks are used to apply AC and DC stress, the inner blue stack is used to apply further DC offset stress. The cell body temperature is measured using a Cernox 1050 temperature sensor. (c) Photograph of a sample mounted on a CS100 cell for an elastocaloric effect measurement. The crystallographic [100] axis is oriented along the uniaxial stress direction $\sigma_{\rm xx}$.}
 \label{Fig1}
\end{figure}

Our measurements are performed using the commercially available uniaxial stress apparatus CS100 from \textit{Razorbill instruments}. The CS100 cell is designed to compensate for the thermal expansion of the lead zirconate titanate (PZT) stacks \cite{Hic14.1}. Furthermore, due to matching of the thermal expansion of Ba(Fe$_{1-x}$Co$_x$)$_2$As$_2$ (Ref.\,\onlinecite{Dal09.1}) and the sample mounting plates (titanium) \cite{Ike18.1}, the externally applied strain on the sample is almost perfectly independent of temperature for a fixed voltage applied to the PZT stacks. Uniaxial stress is applied along a bar shaped sample of Ba(Fe$_{0.975}$Co$_{0.025}$)$_2$As$_2$ (with dimension 2200$\rm\mu$m x 550$\rm\mu$m x 35$\rm\mu$m) by affixing it in between two mounting plates that are pushed together/pulled apart using voltage controlled PZT stacks (see Fig.\,\ref{Fig1}). All single crystals studied here were grown using a FeAs self flux technique described elsewhere\cite{Chu09.1}. For the uniaxial stress experiment the crystallographic [100] direction is aligned along the direction of the stress $\sigma_{\rm xx}$. While the inner PZT stack is used to apply constant DC offset stress, the outer stacks (red stacks in Fig.\,\ref{Fig1}(b)) are used in parallel to apply oscillating stress optionally on top of additional DC offset stress. The outer stacks are controlled by the oscillator output voltage of a \textit{Stanford Research} SR860 Lock-In amplifier (whose input is used for detecting the elastocaloric temperature oscillation) amplified by a factor of 25 by a circuit using a \textit{Tegam} 2350 precision power amplifier. The inner stack is controlled by a DC voltage supplied by an auxiliary output of the SR860 amplified by a factor of 25 using a \textit{piezosystem jena} SVR 350-1 bipolar voltage amplifier. The CS100 cell is equipped with a capacitive displacement sensor, measuring the distance change between the mounting plates. The externally applied nominal strain in the sample along the stress axis ($\varepsilon_{\rm xx}^{\rm disp}$) is then determined by scaling the measured length change by the zero volt length of the sample between the mounting plates. The strain transfer ratio $\varepsilon_{\rm xx}/\varepsilon_{\rm xx}^{\rm disp}$ can be estimated using finite element simulations and has been reported \cite{Ike18.1} to be 0.7$\pm$0.07 for a setup (similar sample and mounting technique) comparable to the one investigated here. 

The capacitive sensor is sampled using an \textit{Andeen-Hagerling} AH2550A capacitance bridge.
The sampling rate of this bridge is not fast enough to allow for the direct measurement of the AC strain amplitude. To estimate this quantity, we first measure the displacement per volt of the outer stacks in DC mode throughout the temperature window of interest (see Appendix\,\ref{sec:Exp}, Fig.\,\ref{FigA1}). The stroke length of PZT stacks generally also depends on the frequency. It has, however, been shown \cite{Hri18.1} for similar PZT stacks that the frequency dependence of the stroke length is small below about 3\,kHz. Instead of application of DC offset stress, the inner PZT stack can also be used to measure a proxy for the dynamic stress applied by the outer PZT stacks\cite{Din18.1}. We use this technique to determine the change of the stroke length of the outer stacks as a function of frequency and correct our data accordingly (for details see Appendix\,\ref{sec:Exp:stroke}). The oscillating strain amplitude is then estimated from the driving AC voltage and the measured displacement per volt. The maximum strain frequency tested here is about 950\,Hz. Fixed frequency temperature sweeps are performed using strains driven by an 10V peak-to-peak AC voltage at a frequency of 21\,Hz.

The elastocaloric temperature oscillation is measured using a type-E thermocouple\footnote{To convert the measured voltage oscillation into a temperature oscillation the NIST ITS-90 table for type E thermocouples is used as calibration.}, consisting of Chromel and Constantan thermocouple wire with a diameter of 12.5\,$\mu$m. The thermocouple is attached to the sample using a thin layer of AngstromBond AB9110LV. The thermocouple is referenced to the cell body temperature by soldering its ends onto thermally heat sunk pads on the cell body (see Figure\,\ref{Fig1}(b)). The thermocouple voltage is first amplified using a Stanford Research SR554 transformer preamplifier in passive mode followed by Stanford Research SR560 set to a gain of 200. The input filter is configured as a second order high-pass filter with characteristic frequency of 0.3\,Hz. The transfer function of the preamplifier chain is measured by exciting a small voltage of known amplitude and varying frequency across the thermocouple and measuring the preamplified voltage amplitude and phase. As the transfer function of the transformer preamplifier depends on the thermocouple resistance, this measurement is repeated at several temperatures across the temperature window of interest. All elastocaloric effect data presented below is corrected for this effect. The preamplified signal is detected using a Stanford Research SR860 Lock-In amplifier. For fixed frequency temperature sweeps, the SR860 time constant is set to 3s (unless stated otherwise) such that for the applied filter settings (24dB advanced) and taking into account a sweep rate of 0.5K/min, the data is effectively correlated across roughly 150\,mK. The average sample temperature is measured using a Cernox CX-1050 temperature sensor from Lakeshore mounted on the Ti body of the CS100 cell. The temperature sensor is sampled using a Lakeshore 340 temperature controller.

\subsection{Thermodynamic conditions of the experiment}
The elastocaloric coefficient discussed above (Eqns.\,\ref{Eq1} and \ref{ECCP}) has been derived under adiabatic conditions ($dS=0$) and starting from the Gibbs free energy, which is determined by its state variables - the stress components (note that for this discussion, we neglect the small AC strain/stress perturbations and consider only the DC offset values) as well as temperature. As such, the partial derivative of entropy with respect to temperature is taken at constant stress which is reflected by the heat capacity at constant stress in Eqns.\,\ref{Eq1} and \ref{ECCP}. For the samples investigated below, due to the thermal expansion matching of the sample material and titanium (cell material), the strain as a function of temperature is, for zero externally applied strain, approximately fixed to the value of the freestanding material. If the relevant elastic moduli are temperature independent, this implies that the stress applied to the sample using the PZT stacks is held constant under changing temperatures. If, however, the elastic moduli are strongly temperature dependent, the externally applied stress is no longer independent of temperature (even for a sample with its thermal expansion perfectly matched to the expansion of titanium). Instead, in this case only the externally applied strain along the stress direction (as well as all other contributions of externally applied strains to the strain tensor) are kept constant under changing temperatures. While these more complicated conditions could in principle be treated, the heat capacity under these conditions can does not differ appreciably from the heat capacity under constant stress. Indeed  the heat capacity at constant strain $C_{\varepsilon}$ differs from $C_\sigma$ only by a small correction. A more detailed discussion is given in Appendix\,\ref{sec:thermodyn}. For the specific experiments discussed below, the relevant elastic moduli [for uniaxial stress along the tetragonal [100] axis of Co doped BaFe$_2$As$_2$ ($c_{\rm B1g}$, $c_{\rm A1g,1}$, and $c_{\rm A1g,2}$)] are roughly temperature independent\cite{Yos12.1}.

\subsection{Discretized thermal modelling}
\begin{figure}
 \includegraphics[width=0.5\textwidth]{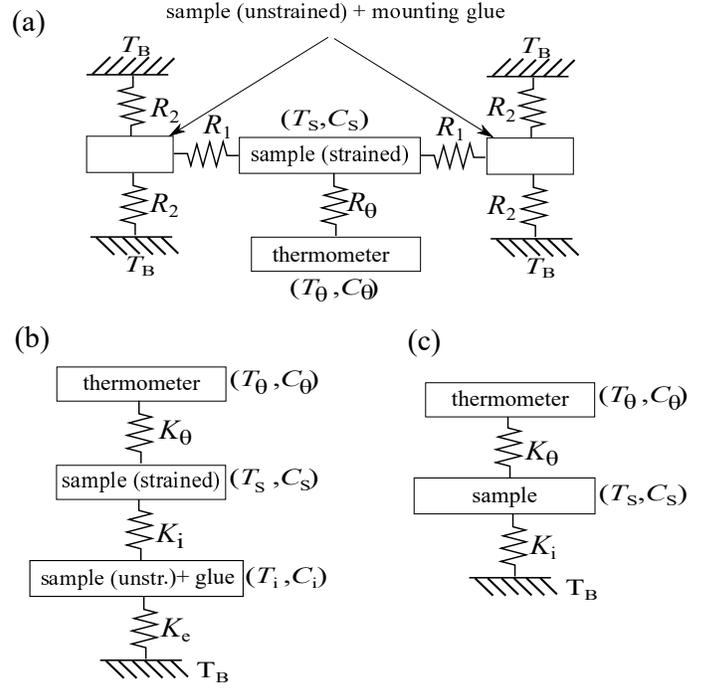}
 \caption{(a) Discretized thermal model of the elastocaloric effect experiment considered here. The strained parts of the sample between the mounting plates experience an AC temperature oscillation that leads to heat flow through the thermal resistances $R_1$ to and from the unstrained parts of the sample. These parts are further connected to the thermal bath at temperature $T_{\rm B}$ vie thermal resistances $R_2$. The thermometer has finite heat capacity and is connected to the sample vie the thermal resistance $R_{\theta}$. (b) Reduced equivalent sketch of the model shown in (a). (c) Simplified model absorbing the unstrained parts of the sample into the thermal bath.}
 \label{Fig2}
\end{figure}

Discretized thermal modelling is well established for AC calorimetry\cite{Sul68.1,Vel92.1,Rio04.1} and provides good intuition and semi quantitative understanding of thermal experiments. Figure\,\ref{Fig2}(a) shows the full discretized thermal model for our experiment described above. This full model is simplified to the equivalent model shown in Figure\,\ref{Fig2}(b). A thermometer of heat capacity $C_{\rm \theta}$ is connected to the strained parts of the sample (with a heat capacity $C_{\rm S}$) through a path of thermal conductance $K_{\rm \theta}$. Due to the elastocaloric effect, the temperature of the strained part of the sample (between the mounting plates) generates elastocaloric temperature oscillation which is described by the time dependence of the sample temperature $T_{\rm S}(t)$. The temperature gradient towards the unstrained parts of the sample (with heat capacity $C_{\rm i}$, which also includes contributions from the mounting glue) yields heat flow through a path of conductance $K_{\rm i}$. The unstrained parts of the sample are connected to the thermal bath at temperature $T_{\rm B}$ through the mounting glue. Thus, an additional heat path $K_{\rm e}$ is considered. The internal time constant of thermalization $\tau_{\rm i}=C_{\rm S}/K_{\rm i}$ sets the characteristic time scale for thermalization of the strained and the unstrained parts of the sample. The external time constant $\tau_{\rm e}=C_{\rm i}/K_{\rm e}$ describes the thermalization to the bath. Finally, the thermometer time constant $\tau_{\theta}=C_{\theta}/K_{\theta}$ sets the time scale of thermalization of the thermometer to the sample. Neglecting the unstrained parts of the sample results in the even simpler thermal model shown in Fig.\,\ref{Fig2}(c). As shown below, quantitatively, the two simplified models differ only by a very small amount for parameters appropriate for the experiment discussed below. We thus start by discussing the simplest model, for which the heat flow equations read:
\begin{equation}
\begin{aligned}
&\dot{T_{\rm S}} =E_0 \dot{\varepsilon}_{\rm xx}-(T_{\rm S}-T_{\rm B})\frac{K_{\rm i}}{C_{\rm S}}\\
&\hspace*{1.5cm}-(T_{\rm S}-T_{\rm \theta})\frac{K_{\rm \theta}}{C_{\rm S}},\\
& \dot{T_{\rm \theta}} =-(T_{\rm \theta}-T_{\rm S})\frac{K_{\rm \theta}}{C_{\rm \theta}}.
\end{aligned}
\end{equation}
The elastocaloric effect is introduced by considering $T_{\rm S}(t)=T_{\rm S0}+E_0 \varepsilon_{\rm xx}(t)$, with $E_0=dT/d\varepsilon_{\rm xx}$, and $\varepsilon_{\rm xx}(t)=\varepsilon_{\rm xx,0}\sin(\omega t)$. In the absence of the elastocaloric effect, all temperatures exponentially tend towards $T_{\rm B}$. Focusing on the particular solution of the second order differential equation for $T_{\rm \theta}$ we find:

\begin{equation}
\begin{aligned}
T_{\rm \theta}(t) &=T_{\rm B}+\frac{E_0\cdot \varepsilon_{\rm xx,0}}{\sqrt{a^2+b^2}}\sin(\omega t +\phi),\\
\phi &=\arctan\left(\frac{a}{b}\right),\hspace{0.5cm}\text{with}\\
a &=\frac{1}{\omega \tau_{\rm i}}-\omega\tau_{\rm \theta},\hspace{0.5cm}\text{and}\\
b &=1+\frac{C_{\rm \theta}}{C_{\rm S}}+\frac{\tau_{\rm \theta}}{\tau_{\rm i}}.
\label{eq:solution1}
\end{aligned}
\end{equation}
The solution for the more complicated model is found in a similar way and discussed in Appendix\,\ref{sec:TFs}. 

\begin{figure}
 \includegraphics[width=0.49\textwidth]{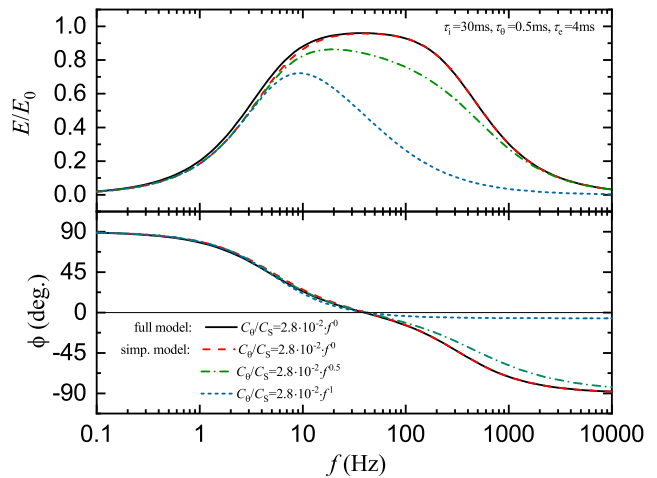}
 \caption{Thermal transfer function for the fully discretized model (Fig.\,\ref{Fig2}(b)) (black lines), as well as the simplified model neglecting the unstraind parts of the sample (Fig.\,\ref{Fig2}(c)) (red dashed line). The top panel shows the measured amplitude of the elastocaloric temperature oscillation rescaled by the intrinsically expected value $E_0$. The lower panel shows the phase of the elastocaloric temperature oscillation. Considering heat capacities of $C_{\rm i}=2.4\times10^{-5}$\,J/K, $C_{\rm S}=1.7\times10^{-5}$\,J/K, and $C_{\rm \theta}=4.7\times10^{-5}$\,J/K, as well as time constants of thermalization of $\tau_{\rm i}=30$\,ms, $\tau_{\rm e}=4$\,ms, and  $\tau_{\rm \theta}=0.5$\,ms, both models yield only slightly different results. The ratio $\frac{C_{\rm \theta}}{C_{\rm i}}$ is expected to be frequency dependent. The dash-dotted green, as well as the dotted blue line show the transfer function, altered, considering a correction according to 1D and 2D heat flow, respectively.}
 \label{Fig3}
\end{figure}

    Figure\,\ref{Fig3} shows the thermal transfer function (measured amplitude $E=E_{0}/\sqrt{\left(a^2+b^2\right)}$ normalized by the nominally expected amplitude of the elastocaloric temperature oscillation $E_{0}$ (top panel), and phase $\phi$ of $T_{\rm \theta}(t)$ (bottom panel) for a set of parameters, estimated for our experiment. As mentioned above, we find that the unstrained parts of the sample affect the thermal behavior of the experiment only very slightly (compare black and dashed red lines in Fig.\,\ref{Fig3}). In order for elastocaloric effect measurements to result in absolute units, the experiment is ideally set up in a way such that the thermal transfer function shows a wide plateau in $E/E_{0}$ at a value of 1. In such a scenario the phase of the elastocaloric temperature oscillation levels off around zero degrees (or 180 degrees dependent on the sign of the ECE) in the frequency region around the amplitude plateaus. Examining the solution for the elastocaloric temperature oscillation (Eq.\ref{eq:solution1}) it can be seen that the width of the plateau is determined by the two time constants, that of the thermometer and the internal thermalization of the sample $\tau_{\rm \theta}$ and $\tau_{\rm i}$, respectively.  Ideally $\tau_{\rm i}>100\cdot\tau_{\rm \theta}$ for a reasonably wide plateau. The maximum value of $E/E_{0}$ is given by $1/b=(1+C_{\rm \theta}/C_{\rm S}+\tau_{\rm \theta}/\tau_{\rm i})^{-1}$. Thus, in addition to a small ratio $\tau_{\rm \theta}/\tau_{\rm i}$, a small ratio of heat capacities $C_{\rm \theta}/C_{\rm S}$ is required so the maximum of $E/E_{0}$ is roughly 1.
    
    One important point neglected within this thermal model is the frequency dependence of $C_{\rm \theta}/C_{\rm S}$ resulting from the fact that the characteristic thermal length within the sample (and thus the volume of the sample thermalizing with the thermometer) is frequency dependent. As long as the characteristic thermal length (proportional to $\sqrt{2D\omega^{-1}}$, where $D$ abbreviates the thermal diffusivity) is larger than the width and the thickness of the sample, 1D heat flow is considered leading to a correction of the form $\frac{C_{\rm \theta}}{C_{\rm i}}\propto \omega^{0.5}$. In case the characteristic thermal length is shorter than the width of the sample, the resulting 2D heat flow problem causes a correction of the form $\frac{C_{\rm \theta}}{C_{\rm i}}\propto \omega$. As the strain frequencies studied here do not allow for accessing characteristic thermal length on the order of the sample thickness, the 3D case is not considered. As shown in Fig.\,\ref{Fig3}, the 1D case (dash-dotted green line) causes the amplitude part of the transfer function to become asymmetric (when plotted versus logarithmic frequency), but changes the phase only slightly. The 2D case (dotted blue line in Fig.\,\ref{Fig3}) results in a largely suppressed plateau in $E/E_{0}$ and a phase that shows only weak frequency dependence at high frequencies. For a thermal diffusivity of $7.4\times10^{-6}$\,m$^2$s$^{-1}$ appropriate for the materials\footnote{The thermal diffusivity is estimated for a temperature of 100K from a thermal conductivity of $\kappa=10$\,Wm$^{-1}$K$^{-1}$ Ref.\onlinecite{Mac09.1}, the measured heat capacity of 210\,J kg$^{-1}$K$^{-1}$ and a density of 6400\,kg m$^{-3}$.} studied here at a temperature of 100\,K, the heat capacity ratio should be expected to become frequency dependent above frequencies of order 10\,Hz. Furthermore a crossover from 1D to 2D at frequencies of order 50-100\,Hz is expected. 

In case $E/E_{0}$ does not show a plateau due to a large $\tau_{\rm \theta}/\tau_{\rm i}$ or a frequency dependence of $C_{\rm \theta}/C_{\rm S}$, the temperature dependence of the elastocaloric effect can still be measured correctly. In this case, however, the temperature dependence of the thermal transfer function has to be determined carefully so that the measured elastocaloric temperature oscillation can be corrected for this effect. In practice, the most effective strategy to achieve acceptable absolute accuracy is to minimize the heat capacity of the thermometer. This reduces $C_{\rm \theta}/C_{\rm S}$ and at the same time decreases $\tau_{\rm \theta}$. Here, we use the tip of a thermocouple made out of 12.5\,$\rm\mu$m thermocouple wires as our thermometer. For perfect thermal behavior, a thinfilm thermocouple could be deposited onto an electrically insulated sample surface.

\subsection{Experimental error sources}
The elastocaloric temperature oscillations of the sample are measured using a single ended thermocouple, referenced on the Razorbill CS100 cell body (see Fig.\,\ref{Fig1}). The cell body thus has to be checked for temperature oscillations unrelated to the elastocaloric effect of the sample. Such oscillations could in principle result from the elastocaloric effect of the PZT stacks. Note that electrical losses within the PZT stacks would lead to temperature oscillations at double the drive frequency and thus can not affect our measurement (this is a further advantage of the AC-ECE technique). The contact pads of the measurement thermocouple are checked for spurious AC temperature oscillations using a second (control) thermocouple referenced at a large Cu block used for thermal anchoring the electrical leads of the CS100 cell. As shown in Appendix\,\ref{sec:Exp} (Fig.\,\ref{FigA2}), no spurious temperature oscillations are resolved against the noise background at the strain frequency used for temperature sweeps. With AC temperature oscillations originating from elastocaloric effect within the PZT stacks excluded, the cell body is also checked for DC heating from power dissipation within the stacks. This effect would pose an error source for the average sample temperature measurement made using the Cernox-1050 temperature sensor attached to the cell body. To check for DC heating, we measure (i) the DC voltage across the control thermocouple which measures the temperature difference of the contact pads with respect to a heat sink on the probe, and (ii) the Cernox-1050 sensor installed on the cell body. To differentiate a real heating effect from slow temperature drifts, we toggle the AC voltage excitation to the outer PZT stacks in 30\,s intervals. As is shown in Fig.\,\ref{FigA3} neither the thermocouple nor the Cernox sensor show any sign of DC heating resulting from losses within the PZT stacks. As DC heating is expected to increase with frequency, this effect was checked at a frequency much larger (512\,Hz) than the frequency used within temperature sweeps (21\,Hz). 

\section{Demonstration of an elastocaloric effect measurement on an underdoped iron pnictide} 
Below we present our experiments on the iron based superconductor Ba(Fe$_{0.979}$Co$_{0.021}$)$_2$As$_2$. This material undergoes two subsequent phase transitions\cite{Chu09.1} upon cooling. First, the material experiences an electronically driven coupled nematic/structural phase transition at $T_{\rm S}=$105.3\,K, that is characterized by an electronic nematic order parameter with $B_{\rm 2g}$ symmetry and an associated structural distortion (shear strain $\varepsilon_{\rm B2g}\equiv\varepsilon_{\rm xy}$) that lowers the crystal symmetry from tetragonal to orthorombic. This transition is followed by an antiferromagnetic transition at $T_{\rm N}=$100.5\,K into a state characterized by stripe magnetic order. We study these two second order phase transitions by measuring the elastocaloric effect induced by an oscillating stress $\sigma_{\rm xx}$ oriented along the tetragonal [100] axis, which causes strains $\varepsilon_{\rm xx}$, $\varepsilon_{\rm yy}$, and $\varepsilon_{\rm zz}$. It is most insightful to represent the strain field in the sample in terms of the irreducible representations of the crystallographic point group. For our experiment, we expect the strain components $\varepsilon_{\rm A1g,1}=1/2\cdot(\varepsilon_{\rm xx}+\varepsilon_{\rm yy})$, $\varepsilon_{\rm A1g,2}=\varepsilon_{\rm zz}$, and $\varepsilon_{\rm B1g}=1/2\cdot(\varepsilon_{\rm xx}-\varepsilon_{\rm yy})$ (see Ref.\,\onlinecite{Ike18.1}). As these strains do not induce a finite nematic order parameter\cite{Ike18.1}, both the nematic as well as the antiferromagnetic phase transitions are expected to persist even in the presence of finite stress along [100]. Since B$_{1g}$ strain breaks the C4 rotational symmetry of the tetragonal lattice, its effect on $T_{\rm S}$ is quadratic to leading order\cite{Ike18.1}. Therefore, around zero DC offset stress, the elastocaloric effect due to $\varepsilon_{\rm B1g}$ is small and occurs at double the strain frequency (2$\omega$). The symmetry preserving strain components $\varepsilon_{\rm A1g,1}$ and $\varepsilon_{\rm A1g,2}$ on the other hand yield an ECE at $\omega$ which we observe in the experiments discussed below.

\begin{figure}
 \includegraphics[width=0.5\textwidth]{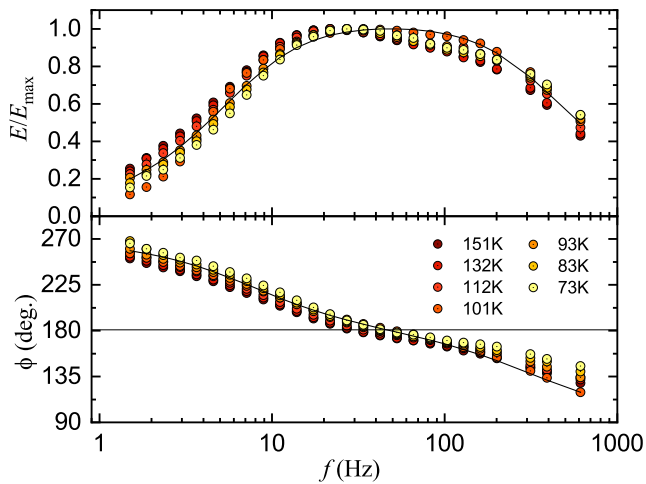}
 \caption{(top) Normalized amplitude of the elastocaloric temperature oscillation measured at different temperatures throughout the temperature window of interest versus strain frequency $f$. (Bottom) The phase of the measured elastocaloric temperature oscillation. The black lines correspond to a fit to the data at 93\,K. Low and high frequency data is well described by the fit, allowing to extract estimates for $\tau_{\rm \theta}=5\cdot10^{-4}$\,s and $\tau_{\rm i}=2.1\cdot10^{-2}$\,s.}
 \label{Fig4}
\end{figure}

Figure\,\ref{Fig4} shows the measured thermal transfer function of the ECE experiment on Ba(Fe$_{0.979}$Co$_{0.021}$)$_2$As$_2$ for temperatures around the nematic and antiferromagnetic transition. Qualitatively, the transfer functions can be well understood using intuition from the discretized thermal model discussed above. At low frequencies, the characteristic thermal length is larger than half the distance between the mounting plates, allowing heat flow in and out of the sample.  Increasing the frequency improves the adiabatic condition until a plateau in the transfer function is reached. The measured phase decreases from 270$^\circ$ to reach 180$^\circ$ (the sample exhibits EC \emph{cooling} upon tensile strains since $dT_{\rm S}/d\varepsilon_{\rm xx}<0$) within the plateau region. The absolute magnitude of the plateau is given by $b^{-1}$ (see Eq.\ref{eq:solution1}, $a\approx1$ in the plateau region) and is thus determined by the heat capacity ratio of thermometer and sample and the ratio of time constants of thermalization $\tau_{\rm \theta}/\tau_{\rm i}$ . Since the heat capacity ratio increases with frequency ($C_{\rm \theta}/C_{\rm S}\propto\omega^{0.5}$ at intermediate frequencies) the measured elastocaloric amplitude decreases even in the region of the plateau $(\tau_{\rm \theta}^{-1}>f>\tau_{\rm i}^{-1}$). This effect is weakest for the data taken at 101\,K, as this temperature is close to the antiferromagnetic transition temperature of the material where the sample heat capacity is enhanced by about 15\%. Further increasing the frequency to values $f>\tau_{\rm \theta}^{-1}$ keeps the sample in a quasi adiabatic regime, but the EC temperature oscillation is progressively decoupled from the thermometer at these higher frequencies.

A quantitative analysis is difficult, since the exact functional form of $C_{\rm \theta}/C_{\rm S}(\omega)$ is unknown. In practice, decreasing the thermometer heat capacity e.g. by using a thinfilm thermocouple would restore a perfect plateau and thus enable the exact measurement of the absolute ECE. Here, we are mainly interested in the temperature dependence of the ECE which carries the most essential information on the phase transitions under investigation.

\begin{figure}
 \includegraphics[width=0.5\textwidth]{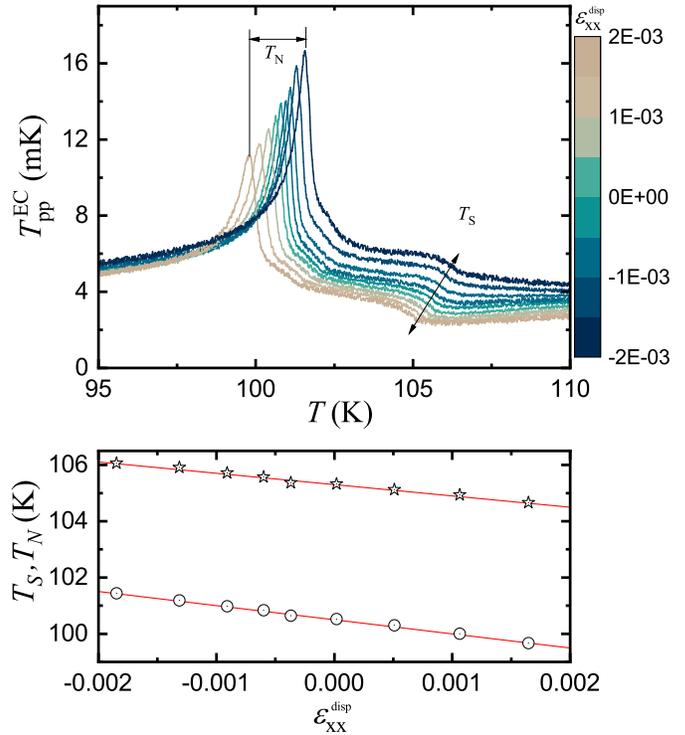}
 \caption{(top) Peak to peak amplitude of the EC temperature oscillation measured during fixed frequency temperature sweeps versus temperature. The data shown was taken during warming runs at a sweep rate of 1\,K/min. The Stanford Research SR860 Lock In Input filter was configured to advanced mode and using a time constant of 1\,s. The bottom panel shows the average transition temperatures determined from cooling and warming data versus nominal strain $\varepsilon_{\rm xx}^{\rm disp}$. The red lines are linear fits to the data.}
 \label{Fig5}
\end{figure}

\begin{figure}
 \includegraphics[width=0.5\textwidth]{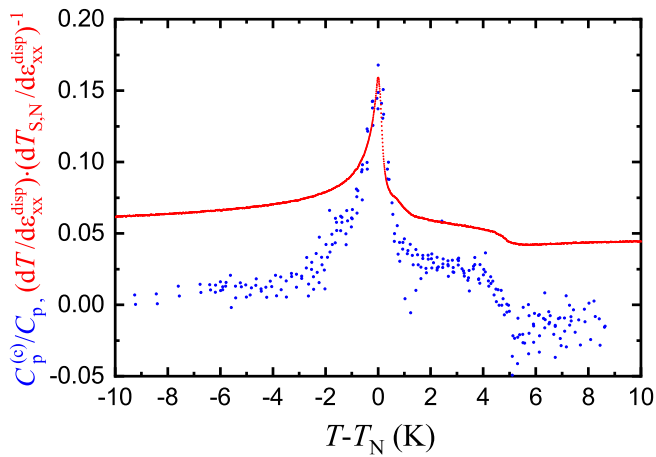}
 \caption{Normalized critical part of the heat capacity (blue dots) around the structural and antiferromagnetic phase transition versus relative temperature $T-T_{\rm N}$. The red symbols show the measured elastocaloric coefficient $dT/d\varepsilon_{\rm xx}^{\rm disp}$ normalized by the average strain dependence of $T_{\rm S}$ and $T_{\rm N}$. The ECE data shown here was detected during a warming run at 0.5\,K/min. The input filter of the detecting SR860 was configured to advanced mode and using a time constant of 3\,s.}
 \label{Fig6}
\end{figure}

In the following paragraph, we discuss errors altering the measured temperature dependence of the ECE. The two main error sources are given by potential temperature dependencies of  $C_{\rm \theta}/C_{\rm S}$ or $\tau_{\rm \theta}/\tau_{\rm i}$. We first turn to the heat capacity ratio $C_{\rm \theta}/C_{\rm S}$. Considering a diffusivity of $7.8\times10^{-6}$\,m$^2$s$^{-1}$, at 21\,Hz (frequency used for temperature sweeps shown in Figs.\,\ref{Fig5}, \ref{Fig6}), approximately 70\% of the sample between the mounting plates thermalize with the thermocouple. The thermometer heat capacity is estimated by the wire contribution (200\,$\rm\mu$m length for each wire) as well as a contribution from the mounting glue layer ($200\times100\times10\rm\mu$m$^3$). Considering appropriate material's properties at 110\,K ($C^{\rm glue}$=1.2\,Jcm$^{-3}$K$^{-1}$, $C^{\rm wire}$=2.2\,Jcm$^{-3}$K$^{-1}$, $C^{\rm sample}$=1.2\,Jcm$^{-3}$K$^{-1}$), we estimate $C_{\rm \theta}/C_{\rm S}(21\,\rm Hz)=0.0155$. A change of the sample's heat capacity within the temperature window of interest of up to 15\% (see Fig.\,\ref{Fig6}) thus has only a very small effect ($<1\%$) on $b$. A second effect introducing extrinsic temperature dependence into the measured ECE is a changing ratio of time constants of thermalization $\tau_{\rm \theta}/\tau_{\rm i}$. To estimate this ratio, we fit the transfer function (amplitude and phase simultaneously) at 93\,K using Eq.\,\ref{eq:solution1}. The cross over from a constant (characteristic thermal length larger than half the distance between the mounting plates) to a frequency dependent heat capacity ratio $C_{\rm r}(f)=C_{\rm \theta}/C_{\rm S}$ is modeled according to $C_{\rm r}(f)=C_{r0} \cdot f^{ \Theta(f-10)\cdot0.5}$, where $\Theta(f-10)$ is a Heaviside function and $C_{r0}=3.4\cdot10^{-3}$ (so that $C_{\rm r}=0.0155$ at 21Hz). The two parameter fit (black lines in Fig.\,\ref{Fig4}) results $\tau_{\rm \theta}=5\cdot10^{-4}$\,s and $\tau_{\rm i}=2.1\cdot10^{-2}$\,s.  
For a frequency of 21\,Hz (used for the fixed frequency temperature sweeps), we thus estimate $b=1.039$. 
Errors in the measured temperature dependence of the ECE due to changing thermal behavior are thus estimated to be small ($<5$\% even if both, $C_{\rm \theta}/C_{\rm S}$ and $\tau_{\rm \theta}/\tau_{\rm i}$ were changing by 100\% within the temperature window of interest). 

The top panel in Fig.\,\ref{Fig5} shows fixed frequency temperature sweeps taken at different DC offset strains $\varepsilon_{\rm xx}^{\rm disp}$. As discussed above, the ECE closely resembles the features observed in heat capacity and thus allows for extracting the critical temperatures of the phase transitions under offset strain. We determine the transition temperatures $T_{\rm S}$ and $T_{\rm N}$ from cooling and warming data and report the average values in the bottom panel of Fig.\ref{Fig5}. Despite the experiment being performed under high vacuum, a relatively small thermal lag of the sample with respect to the CS100 cell body of about 0.1\,K is observed. From the linear fits shown as red lines in the bottom panel of Fig.\,\ref{Fig5}, we determine the strain dependence of the transition temperatures as $dT_{\rm S}/d\varepsilon_{\rm xx}^{\rm disp}=-400$\,K and $dT_{\rm N}/d\varepsilon_{\rm xx}^{\rm disp}=-500$\,K, respectively.

Using the averaged strain dependence of the critical temperatures $dT_{\rm S,N}/d\varepsilon_{\rm xx}^{\rm disp}=0.5\cdot(dT_{\rm S}/d\varepsilon_{\rm xx}^{\rm disp}+dT_{\rm N}/d\varepsilon_{\rm xx}^{\rm disp})=$-450\,K, the elastocaloric temperature oscillation can be, as captured by Eqn.\,\ref{ECCP}, compared to $C_{\rm p}^{\rm (c)}/C_{\rm p}$. This is shown in Fig.\,\ref{Fig6}. The critical part of the heat capacity of Ba(Fe$_{0.979}$Co$_{0.021}$)$_2$As$_2$ is determined by subtracting the measured heat capacity of an overdoped sample with composition Ba(Fe$_{0.89}$Co$_{0.11}$)$_2$As$_2$ that does not undergo a structural or magnetic transition, but is expected to have a very similar non-critical contribution to the heat capacity. The $C_{\rm p}$ measurements on both materials have been performed using the  standard relaxation time calorimetry option of the PPMS from \textit{Quantum Design}. To test for artificial smearing of the heat capacity features, both 1 and 2\% of the sample temperature were used as the set temperature rise for the Ba(Fe$_{0.979}$Co$_{0.021}$)$_2$As$_2$ crystal. Both data sets show a very similar $T$-dependence. Due to the relatively large uncertainty in the sample mass, before subtraction the measured $C_{\rm p}$ of Ba(Fe$_{0.89}$Co$_{0.11}$)$_2$As$_2$ is re-scaled to match the heat capacity of Ba(Fe$_{0.979}$Co$_{0.021}$)$_2$As$_2$ at high temperatures. The blue data in Fig.\,\ref{Fig6} shows the normalized critical contribution $C_{\rm p}^{\rm (c)}/C_{\rm p}$ of Ba(Fe$_{0.979}$Co$_{0.021}$)$_2$As$_2$. The red symbols show the measured elastocaloric coefficient $dT/d\varepsilon_{\rm xx}^{\rm disp}$ normalized by the average strain dependence $dT_{\rm S,N}/d\varepsilon_{\rm xx}^{\rm disp}$. This data represents the full elastocaloric effect which includes, besides the critical contributions, also a weakly temperature dependent non-critical background from lattice or electronic degrees of freedom. As mentioned above, in contrast to the heat capacity, for the elastocaloric effect, the contributions from the critical degrees of freedom dominate the signal. As is obvious from comparing the data in Fig.\,\ref{Fig6}, measuring signatures of continuous phase transitions can be done with superior signal to noise ratio using ECE measurements. Besides the missing background subtraction for the elastocaloric effect data, the two data sets in Fig.\,\ref{Fig6} also differ by a factor of 1.3. This is likely explained by systematic uncertainty introduced by the background subtraction in $C_{\rm p}$, a suppressed plateau in the thermal transfer function of the ECE experiment ($b=1.039$), and partly due to the application of the averaged strain dependence of the two phase transitions $dT_{\rm S,N}/d\varepsilon_{\rm xx}^{\rm disp}$. As mentioned above, in case absolute accuracy is needed, experimentally the best strategy is to reduce the thermometer heat capacity by about one order of magnitude as compared to the heat capacity of the thermocouple used here. Importantly though, using the setup studied here, the temperature dependence of the ECE is measured with high accuracy. The signal-to-noise ratio is clearly superior to that of the calorimetry data also presented in the same figure.

\section{Conclusion}
The AC-ECE technique discussed in this paper enables the study of the elastocaloric effect of materials with high precision. In contrast to standard EC measurement techniques, AC-ECE measurements can be performed under quasi-adiabatic conditions over extended periods of time. Due to the employed phase sensitive lock-in technique, the signal to noise ratio is orders of magnitude larger than that achievable by regular DC measurement techniques. For materials undergoing continuous phase transitions, we have shown that the elastocaloric effect responding to strains that preserve symmetries broken at the transition, resembles the features observed in heat capacity. These features are measured against a small background even at high temperatures. This improves the ability of the technique discussed here to resolve such features even when measured at high temperatures. The technique further allows for tracking the evolution of the heat capacity features under offset strain. Since the elastocaloric temperature oscillation is an intrinsic response of materials to strain, the frequency range of EC measurements is only limited by the time constant of thermalization of the thermometer. Minimizing the thermometer heat capacity, this in principle allows to extend the plateau in the thermal transfer function to high frequencies, facilitating the measurement of the elastocaloric effect in absolute units across wide temperature windows.

\begin{acknowledgments}
The authors thank E. Rosenberg, and P. Massat for insightful discussions. This work was supported by the Department of Energy, Office of Basic Energy Sciences, under Contract No. DEAC02-76SF00515. M.S.I and P.W were supported in part by the Gordon and Betty Moore Foundations EPiQS Initiative through Grant No. GBMF4414. A.T.H, J.C.P, and M.S were supported by a NSF Graduate Research Fellowship (Grant No. DGE-114747) J.C.P was also supported by a Gabilan Stanford Graduate Fellowship, and a Lieberman Fellowship, J.A.W.S acknowledges support as an ABB Stanford Graduate Fellow.

\end{acknowledgments}

\appendix
\section{Further Experimental Details}
\label{sec:Exp}
\subsection{Frequency dependence of the stroke length of PZT stacks}
\label{sec:Exp:stroke}
\begin{figure}[ht]
 \includegraphics[width=0.49\textwidth]{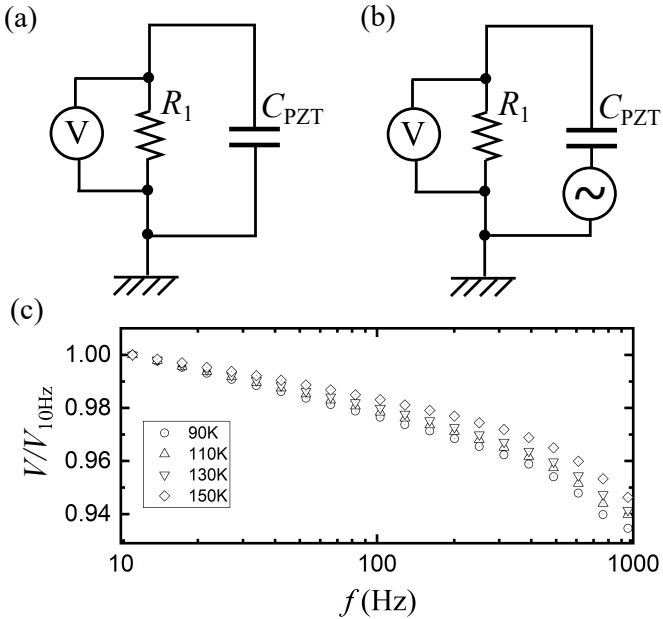}
 \caption{(a) Schematic showing basic working principle of the inner stack as force sensor. A 810\,k$\Omega$ resistor is connected in parallel to the inner stack sketched as the capacitance. (b) Schematic showing how the transfer function of the high-pass filter has been measured. (c) AC Voltage detected across the resistor rescaled by the value measured at the lowest frequency measured. The phase was found to be almost temperature and frequency independent and close to zero.}
 \label{FigA4}
\end{figure}

The change in the stroke length of the outer PZT stacks can be measured by repurposing the inner PZT stack to serve as dynamic force sensor. The stress induced voltage across a piezoelectric material is directly related to the strain in the material. Due to internal charge leakage, PZT stacks can, however, not be used as static force sensors. Dynamic AC force measurements are however possible. Here, frequencies larger than 10\,Hz are investigated. To convert the inner stack into a force sensor, a simple circuit as shown in Fig.\,\ref{FigA4}(a) is used. A 810\,k$\Omega$ resistor is connected in parallel to the inner stack. This resistor prevents a potentially problematic voltage buildup on the PZT stack and at the same time sets the cut off frequency $f_c$ of the high pass filter it forms, together with the capacitance of the stack $C_{PZT}\approx0.5\mu$F, to a relatively small number of about 0.4\,Hz. After correction for the high-pass transfer function, the amplitude and phase of the measured voltage across the resistor is a measure for the force applied to the inner PZT stack. The high pass transfer function is measured by exciting a small AC voltage to one side of the PZT stack as shown in Fig\,\ref{FigA4}(b). Since the Young's modulus of the sample is not expected to show any frequency dependence at these low frequencies, a change in the force applied to the inner PZT stack corresponds to a change in the stroke length of the outer PZT stacks. As shown in Fig.\,\ref{FigA4}(c), we observe a change of less than 10\,\% throughout the frequency window investigated. The thermal transfer functions of our experiment are corrected for this effect accordingly.

\subsection{Estimating the AC strain amplitude}
\begin{figure}[ht]
 \includegraphics[width=0.49\textwidth]{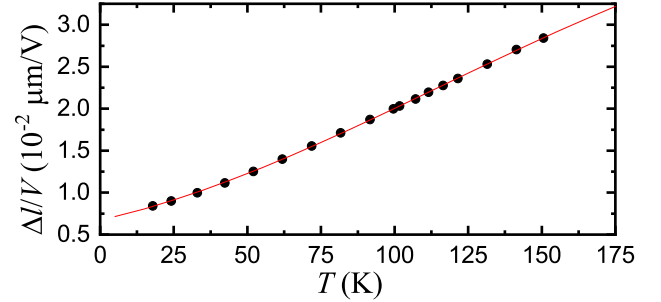}
 \caption{Displacement per volt applied to the outer PZT stacks of the CS100 cell used for the elastocaloric effect measurements as a function of temperature. The red line is polynomial fit to the measured data.}
 \label{FigA1}
\end{figure}

The AC strain amplitude can not be measured by directly sampling the capacitive displacement sensor using our AH2550 capacitance bridge. It can, however, be estimated by first determining the displacement per volt by applying DC voltages to the outer PZT stacks and measuring the corresponding DC displacement. Multiplying this figure by the PZT drive voltage amplitude results the AC strain amplitude. Fig\,\ref{FigA1} shows the displacement per volt for the outer PZT stacks throughout the temperature window of interest. 

\subsection{Spurious temperature oscillations on the CS100 cell body}
\begin{figure}[ht]
 \includegraphics[width=0.49\textwidth]{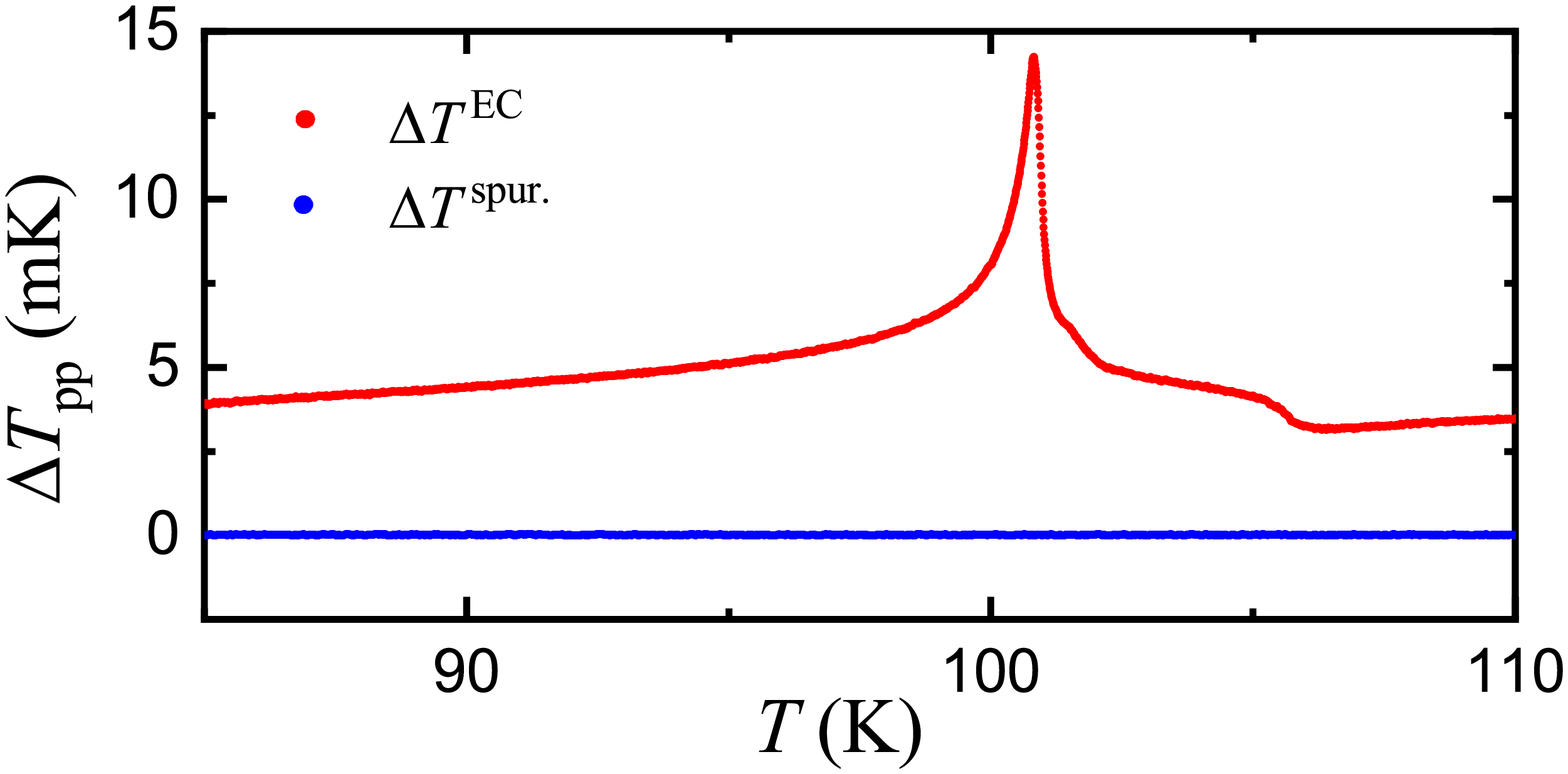}
 \caption{Elastocaloric temperature oscillation of the sample (red symbols) and the temperature oscillations of the contact pads of the measurment thermocouple detected using a second control thermocouple (blue symbols)  versus temperature.}
 \label{FigA2}
\end{figure}

As the CS100 cell body serves as thermal reference for the measurement thermocouple, it has to be carefully checked for spurious temperature oscillations at the measurement frequency. This is done using a second control thermocouple referenced at a large heat sink on the probe. The signal of the control thermocouple is preamplified a Stanford Research SR554 transformer preamplifier in passive mode followed by Stanford Research SR560 set to a gain of 200. The input filter is configured as a second order high-pass filter with characteristic frequency of 0.3\,Hz. No spurious signal is resolved against the noise background which is two orders of magnitude smaller than the elastocaloric signal detected on the sample (see Fig.\,\ref{FigA2}).

\subsection{Effect of power dissipation within the PZT stacks}
\begin{figure}[ht]
 \includegraphics[width=0.49\textwidth]{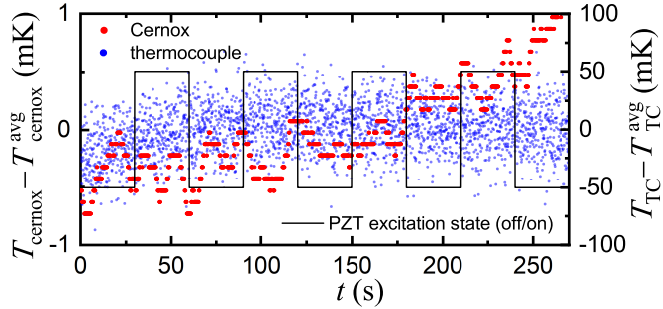}
 \caption{Temperature change of the CS100 cell body versus time. The red symbols show the cell temperature measured using the Cernox sensor, the blue data show the temperature change of the contact pads of the measurement thermocouple measured using the control thermocouple in DC mode. In both cases, the average offset has been subtracted from the data. The voltage excitation (10\,V$_{\rm pp}$, 512\,Hz) to the outer PZT stacks of the CS100 cell is toggled in 30\,s intervals (sketched by the grey line).}
 \label{FigA3}
\end{figure}

Selfheating of PZT stacks from power dissipation has been reported\cite{Hri18.1} to be on the order of serveral Kelvin for freestanding PZT stacks at frequencies of several kHz. The highest frequencies studied here are below 1\,kHz (all fixed frequency temperature sweeps are performed using a strain frequency of 21\,Hz) and moreover, the PZT stacks are connected to a titanium cell of comparatively large heat capacity. A large DC heating effect is thus not expected. Figure\,\ref{FigA3} shows the cell temperature measured using the installed Cernox sensor (red symbols), as well as using the control thermocouple (blue dots) in DC mode. In each case, the average offset has been subtracted from the data. The DC voltage of the control thermocouple is detected without preamplification using a \textit{Keithley} 2000 Multimeter. The measurement has been carried out using an AC voltage excitation on the outer PZT stacks of 10\,V$_{\rm pp}$ at 512\,Hz. The measurement was carried out at the lower limit of the temperature window of interest at 76\,K (where the heat capacity of the cell body is smallest). In order to distinguish a DC selfheating effect from temperature drifts, the voltage excitation to the PZT stacks is toggled in 30\,s intervals. No self heating related temperature change is observed upon toggling the PZT drive voltage at these frequencies.

\section{Thermal transfer functions}
\label{sec:TFs}
The differential equations for the more complete model shown in Fig.\,\ref{Fig2}(b) read:
\begin{equation}
\begin{aligned}
\dot{T_{\rm S}} &=\omega E \cos(\omega t) -\left(T_{\rm S}-T_{\rm i}\right)\frac{K_{\rm i}}{C_{\rm S}}-\left(T_{\rm S}-T_{\rm \theta}\right)\frac{K_{\rm \theta}}{C_{\rm S}},\\
\dot{T_{\rm i}} &=-\left(T_{\rm i}-T_{\rm S}\right)\frac{K_{\rm i}}{C_{\rm i}} -\left(T_{\rm i}-T_{\rm B}\right)\frac{K_{\rm B}}{C_{\rm i}},\\
\dot{T_{\rm \theta}} &=-\left(T_{\rm \theta}-T_{\rm S}\right)\frac{K_{\rm \theta}}{C_{\rm \theta}}.
\end{aligned}
\end{equation}
The elastocaloric effect is again introduced by considering $T_{\rm S}(t)=T_{\rm S0}+E_0 \varepsilon_{\rm xx}(t)$, with $E_0=dT/d\varepsilon_{\rm xx}$, and $\varepsilon_{\rm xx}(t)=\varepsilon_{\rm xx,0}\sin(\omega t)$. In the absence of the elastocaloric effect, all temperatures exponentially tend to $T_{\rm B}$. We thus focus on the particular solution of the third order differential equation for $T_{\rm \theta}$ and find:
\begin{equation}
\begin{aligned}
&T_{\rm \theta}(t) =T_{\rm B}+\frac{E_0 \varepsilon_{\rm xx,0} \sqrt{\left(\frac{C_{\rm S}}{C_{\rm i}}+\frac{\tau_{\rm i}}{\tau_{\rm e}}\right)^2+\tau_{\rm i}^2\omega^2}}{\sqrt{a^2+b^2}}\sin(\omega t +\phi),\\
&\phi =\arctan\left(\frac{b\left(\frac{K_{\rm i}}{C_{\rm i}}+\frac{1}{\tau_{\rm e}}\right)+a \omega}{a\left(\frac{K_{\rm i}}{C_{\rm i}}+\frac{1}{\tau_{\rm e}}\right)+b \omega}\right), \hspace{0.5cm} \text{where}\\
&a = 1+\frac{C_{\rm S}}{C_{\rm i}}+\frac{C_{\rm \theta}}{C_{\rm i}}+\frac{\tau_{\rm i}}{\tau_{\rm e}}+\frac{\tau_{\rm \theta}}{\tau_{\rm i}}+\frac{K_{\rm e}}{K_{\rm i}}\frac{C_{\rm \theta}}{C_{\rm i}}-\tau_{\rm \theta} \tau_{\rm i}\omega^2,\\
&b = \frac{1}{\tau_{\rm e}\omega}-\omega\left(\tau_{\rm i}+\tau_{\rm \theta}+\tau_{\rm \theta} \tau_{\rm i}\frac{K_{\rm \theta}}{C_{\rm S}}+\tau_{\rm \theta} \tau_{\rm i}\frac{K_{\rm i}}{C_{\rm i}}+\frac{\tau_{\rm \theta} \tau_{\rm i}}{\tau_{\rm e}}\right).
\label{eq:solution3}
\end{aligned}
\end{equation}
\section{Thermodynamic considerations}
\label{sec:thermodyn}
Here we derive the relation between the elastocaloric effect and the critical contributions of the heat capacity under uniaxial stress through entropy considerations. We partly follow the derivation discussing the case for hydrostatic pressure reported earlier by Souza et al.\cite{Sou05.1}. We discuss the derivations starting from the Gibbs free energy, where the stress components $\sigma_{\rm ij}$ and $T$ are chosen as the state variables. Finally, we compare the solutions derived from Gibbs and Helmholtz free energy.

Let $T_{\rm c}(\sigma_{\rm kl})$ be phase transition lines in the ($\sigma_{\rm kl}$, $T$) planes. In case $T_{\rm c}(\sigma_{\rm kl})$ is invertible to $\sigma_{\rm kl}^{\rm c}(T)$, $S(T,\sigma_{\rm kl}^{\rm c})$ is also a transition line. When moving in parallel to the transition lines in the ($\sigma_{\rm kl}$, $T$) planes, the corresponding entropy also moves in parallel to the transition line $S(T,\sigma_{\rm kl}^{c})$ [see Figure\,\ref{Fig0}(b) for the closely related dependence of $S$ on temperature and strain $\varepsilon_{\rm xx}$; $\sigma_{\rm xx}=(c_{\rm xxxx}-\nu\cdot c_{\rm xxyy}-\nu^\prime\cdot c_{\rm xxzz})\varepsilon_{\rm xx}$]. The change of entropy is related to the changes $dT$, $d\sigma_{\rm kl}^{\rm c}$ via the total derivative:
  
\begin{equation}
\begin{aligned}
\frac{dS^{\rm (tl)}}{dT}=& \left(\frac{\partial S}{\partial \sigma_{\rm kl}}\right)_{T_{\rm c}+\delta} \frac{d\sigma_{\rm kl}^{\rm c}}{dT}+\left(\frac{\partial S}{\partial T}\right)_{\sigma_{\rm ij}}=\\
=&\alpha_{\rm ij}\frac{d\sigma_{\rm kl}^{\rm c}}{dT}+\frac{C_{\sigma}}{T}.
\label{C1}
\end{aligned}
\end{equation}
The superscript in $S^{\rm (tl)}$ highlights that entropy changes corresponding to movements along the phase transition line (with constant distance to the boundary) are considered. This equation holds for small $\delta$ above and below $T_{\rm c}$.

Using $\frac{d\sigma_{\rm kl}^{c}}{dT}=\left(\frac{dT_{\rm c}}{d\sigma_{\rm kl}}\right)^{-1}$ and rearranging Eqn.\,\ref{C1}, we have:

\begin{equation}
\begin{aligned}
T\left[\frac{C_{\sigma}}{T}-\frac{dS^{\rm (tl)}}{dT}\right]=&-T\alpha_{\rm ij}\left(\frac{dT_{\rm c}}{d\sigma_{\rm kl}}\right)^{-1}=\\
=&-T\alpha_{\rm ij}c_{\rm ijkl}\frac{d\epsilon_{kl}}{dT_{\rm c}}.
\end{aligned}
\end{equation}

Considering uniaxial stress along [100] (all $\sigma_{\rm kl}=0$ except for $\sigma_{\rm xx}$) and using relations between the strain components $\varepsilon_{\rm kl}$ ($\varepsilon_{2}=-\nu \varepsilon_{1}, \varepsilon_{3}=-\nu\prime \varepsilon_{1}, \varepsilon_{4},\varepsilon_{5},\varepsilon_{6}=0$) in Voigt notation (1=xx, 2=yy, 3=zz, 4=yz, 5=xz, 6=xy), we can re-write the EC effect (Eq.\,\ref{Eq1}) as:

\begin{equation}
dT=-T\frac{\alpha_{\rm 1}}{C_{\rm \sigma}}c_{\rm 1j}d\varepsilon_{\rm j}=\frac{T \left[\frac{C_{\sigma}}{T}-\frac{dS^{\rm (tl)}}{dT}\right]}{C_{\rm \sigma}}\frac{dT_{\rm c}}{d\varepsilon_{\rm i}}d\varepsilon_{\rm i},
\end{equation}
with $i$=1, 2, or 3. 

As $dS^{\rm (tl)}/dT$ describes entropy changes in parallel to the phase transition line, it is dominated by non critical degrees of freedom. Thus, $T\cdot(dS^{\rm (tl)}/dT)$ approximates the non critical background of the heat capacity. This is more rigorously shown below. Let the entropy consist of contributions from critical and non critical degrees of freedom $S(T,\sigma_{\rm 1})=S^{\rm (c)} + S^{\rm (nc)}$. Generally changes of the entropy can then be written as:
\begin{equation}
\begin{aligned}
dS=& \frac{\partial S^{\rm (nc)}}{\partial T}dT +\frac{\partial S^{\rm (nc)}}{\partial \sigma_{\rm 1}}d\sigma_{\rm 1} +\\
+& \frac{\partial S^{\rm (c)}}{\partial T}dT +\frac{\partial S^{\rm (c)}}{\partial \sigma_{\rm 1}}d\sigma_{\rm 1}.
\end{aligned}
\end{equation}

Assuming the entropy of the non critical degrees of freedom is stress independent, we find for changes along the phase boundary:
\begin{equation}
dS^{\rm (tl)}= \frac{\partial S^{\rm (nc)}}{\partial T}dT + \frac{\partial S^{\rm (c)}}{\partial T}dT +\frac{\partial S^{\rm (c)}}{\partial \sigma_{\rm 1}^{\rm c}}d\sigma_{\rm 1}^{\rm c}.
\end{equation}

As $\sigma_{\rm 1}^{\rm c}(T)$ is invertible to $T_{\rm c}(\sigma_{\rm 1})$, we can rewrite the entropy change $dS^{\rm c}(T,T_{\rm c})$ along the phase boundary as:
\begin{equation}
dS^{\rm (tl)}= \frac{\partial S^{\rm (nc)}}{\partial T}dT + \frac{\partial S^{\rm (c)}}{\partial T}dT +\frac{\partial S^{\rm (c)}}{\partial T_{\rm c}}dT_{\rm c}.
\end{equation}

Moving along the phase boundary $dT_{\rm c}=dT$, thus:
\begin{equation}
\begin{aligned}
&\frac{dS^{\rm (tl)}}{dT}= \frac{\partial S^{\rm (nc)}}{\partial T} = \frac{C_\sigma^{\rm (nc)}}{T} \hspace{0.5cm} \text{for}\\
&\frac{\partial S^{\rm (c)}}{\partial T} +\frac{\partial S^{\rm (c)}}{\partial T_{\rm c}}=0.
\end{aligned}
\end{equation}

Functions of form $S^{\rm (c)}(T-T_{\rm c})$ satisfy the partial differential equation above. Under these conditions we finally have:
\begin{equation}
\begin{aligned}
dT=-T\frac{\alpha_{\rm 1}}{C_{\rm \sigma}}c_{\rm 1j}d\varepsilon_{\rm j}=&\frac{T \left[\frac{C_{\sigma}}{T}-\frac{C_\sigma^{\rm (nc)}}{T}\right]}{C_{\rm \sigma}}\frac{dT_{\rm c}}{d\varepsilon_{\rm i}}d\varepsilon_{\rm i}=\\
=&\frac{C_{\rm \sigma}^{\rm (c)}}{C_{\rm \sigma}}\frac{dT_{\rm c}}{d\varepsilon_{\rm i}}d\varepsilon_{\rm i}.
\end{aligned}
\end{equation}

Starting from Helmholtz free energy, using the strain components $\varepsilon_{\rm ij}$ and temperature $T$ as state variables, a similar derivation then the one presented above yields:
\begin{equation}
dT=\frac{C_{\rm \epsilon}^{\rm (c)}}{C_{\rm \epsilon}}\frac{dT_{\rm c}}{d\varepsilon_{\rm i}}d\varepsilon_{\rm i}. 
\end{equation}

The main difference is thus found in the thermodynamic conditions keeping in one case stress and in the other case strain fixed. The difference in the heat capacities is given by\cite{Bar70.1}:
\begin{equation}
C_{\rm \epsilon}-C_{\rm \sigma}=-T\alpha_{\rm ij}\alpha_{\rm kl}c_{ijkl}^T.
\label{dCp}
\end{equation}
For solids, the thermal expansion coefficients are typically on the order of $10^{-5}$\,K$^{-1}$. The elastic moduli are on the order of $10^{11}$\,Pa. At 100\,K, the above difference is thus roughly  $10^3-10^4$\,J m$^{-3}$K$^{-1}$ while the heat capacity is on the order of $10^6$\,J m$^{-3}$K$^{-1}$. Thus, the heat capacity of perfectly clamped and a freestanding crystal differs only by a small amount at most cryogenic temperatures.

%

\end{document}